\documentclass[epsf,axodraw]{hep99}
\usepackage{epsf}
\begin{document}
\newcommand{\be}{\begin{equation}}
\newcommand{\ee}{\end{equation}}
\newcommand{\lsim}{{\;\raise0.3ex\hbox{$<$\kern-0.75em\raise-1.1ex
\hbox{$\sim$}}\;}}
\newcommand{\gsim}{{\;\raise0.3ex\hbox{$>$\kern-0.75em\raise-1.1ex
\hbox{$\sim$}}\;}}
\newcommand{\lbl}[1]{\label{eq:#1}}

%\begin{flushright}
% HIP-1999-67/TH
%\end{flushright} 

\title{Production of four jets in LR model$^{\dagger}$}

\author{Jukka Maalampi$^1$ and Nikolai Romanenko$^{1,2}$}
%
% Use of footnote symbols, footnoted material after \maketitle 
%\author{A J Cox$^1$\dag\ and Jim Revill$^2$\ddag}

\address{$^1$  Theoretical Physics 
  Division, Department of Physics,
   University of Helsinki, Finland\\
$^2$  Petersburg Nuclear Physics Institute,
   Gatchina, Russia\\[3pt]
E-mails: {\tt Jukka.Maalampi@helsinki.fi, Nikolai.Romanenko@helsinki.fi}}

\abstract{We consider the reaction $e^-e^- \rightarrow 
q\:q\:\bar q\:\bar q$ as
 a test of lepton number non-conservation 
 in the framework of the left-right-symmetric
  electroweak model. The main contributions to this process
   are due to  Majorana neutrino exchange in $t$-channel 
   and  doubly charged Higgs ($\Delta^{--}$) exchange in $s$-channel
with a pair of right-handed weak bosons ($W_R$)
 as intermediate state. We show that
in a linear $e^-e^-$ collider with the collision energy of  1 TeV (1.5 TeV)
the cross section of this process is 0.01 fb (1 fb),
 and it will, for the anticipated
  luminosity of $10^{35}$  cm$^{-2}$s$^{-1}$,
   be detectable below the $W_R$ threshold. We 
   study the sensitivity of the reaction on the masses 
   of the heavy neutrino, $W_R$ and $\Delta^{--}$.}

\maketitle

% Text of footnotes comes after \maketitle
\fntext{*}{  Supported  by the Academy of Finland under the contract
 40677 and by RFFI grant 98-02-18137. \\
Preprint HIP-1999-67/TH}
%\fntext{2}{E-mail: jim.revill@ioppublishing.co.uk}
%\fntext{\dag}{Here is a footnote.}

\section{Introduction}

\setcounter{equation}{0}
  The electroweak model with the left-right (LR) gauge symmetry
$SU(3)_c \otimes SU(2)_L \otimes SU(2)_R \otimes U(1)_{B-L}$,  
proposed in \cite{LR}, is one of the most popular extensions 
of the Standard Model (SM). It gives a better understanding of parity
violation than the SM and it maintains
 the lepton-quark symmetry  in weak interactions. 
Parity is in it  broken spontaneously, and embedding of the model into the SO(10) grand unified scheme \cite{Ell} can be implemented consistently when the
scale of the discrete LR-symmetry breaking is  more than 1 TeV or so.

Perhaps the most important property of the LR-model is its ability to provide, in terms of the seesaw mechanism
\cite{SS}, a simple and  natural explanation to  the smallness of  the masses of the ordinary neutrinos.
 The recent observation by the SuperKamiokande experiment of the atmospheric neutrino oscillations  \cite{Kam} confirmed that
 at least some of the neutrino species do have a mass, giving an
 additional argument in favour of the LR-symmetric model.

An essential ingredient of the LR-model are the triplet Higgses.
 Their interactions with fermions break the lepton number by two units.  In the literature  different observable lepton number violating processes, including
 doubly charged Higgs production \cite{MaalD2},
 vector-boson pair and triple production for
 elect\-ron-po\-sitron and electro-electron colliders 
  \cite{MaalW2,Gun}, have been investigated.

 In the present talk we will consider the 
   lepton-number violating process

\be
e^-e^-\to q\:q\:\bar q\:\bar q
\label{reaction}\ee
with various quark flavour combinations.
 One would expect to obtain indirect
  evidence of the LR-model via this process well 
  below the threshold of $W_R^{\pm}$. The details
of this study may be found in \cite{MaalRom}.

\section{Numerical Results\label{sec:figs}}

 By means of CompHEP \cite{CompHEP} we have derived  the squared matrix elements for $e^-e^- \rightarrow b\:b\:\bar{t}\:\bar{t} $ and computed the ensuing  cross sections  at the collision energies
$\sqrt s=1$ TeV and $\sqrt s=1.5$ TeV. The results depend on a number of unknown parameters of the LR-model, the most important ones
being the masses of the right-handed boson $W_R$ and doubly charged Higgs-Majoron $\Delta^{--}$.
 We 
consider theory without $W_L-W_R$ mixing and neglect
small effects of the seesaw
mixing. We  restrict ourselves
to the manifestly left-right symmetric case, implying that the left and right-handed interactions have the same coupling strength, i.e. $g_L=g_R$,  and
that the  Kobayashi-Maskawa mixings of the right-handed charged currents
are exactly the same as those of the left-handed ones, in particular $V^R_{tb}= V^L_{tb}\equiv V_{tb}$.
 
In Fig. 1 we show the energy dependence
of the total cross section of the process $e^-e^- \rightarrow
b\:b\:\bar{t}\:\bar{t} $  for various values of
masses of the triplet Higgs $\Delta^{--}$ and the 
right-handed neutrino $\nu_2$. In all the cases
the right-handed boson mass is taken to be $M_{W_R}= 700$ GeV.

\begin{figure}
\begin{center}
\epsfysize=10cm
\epsffile{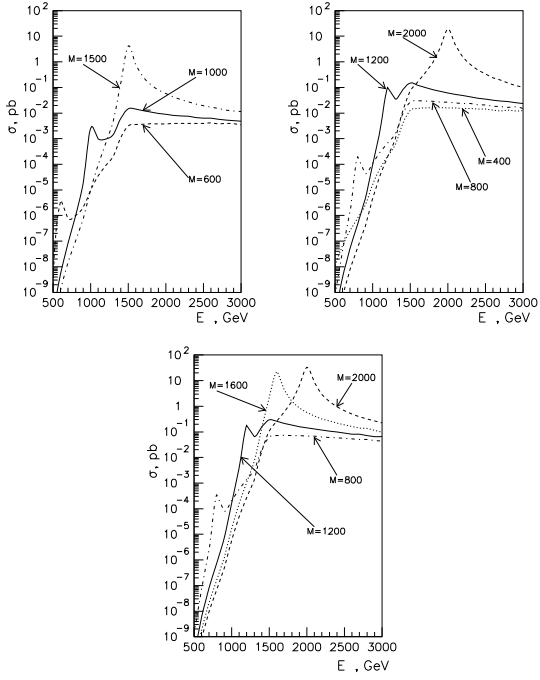}
\end{center}
\caption{Energy dependence
of the full cross section for the process
$e^-e^- \rightarrow b\:b\: \bar{t}\: \bar{t}$ 
  for  different values
 of $\Delta^{--}$ mass
 ($M \equiv M_{\Delta^{--}}$)
 and right-handed neutrino masses:
 $m_{\nu_2}=1$ TeV (left upper picture),
 $m_{\nu_2}=1.5$ TeV (right upper picture),
 $m_{\nu_2}=2$ TeV (lower picture),
 (see comments in the text).}
\label{fe}
\end{figure}

 In Fig. 2
we present the sensitivity contours
for $\sqrt s=1.5$ TeV with the masses of the right-handed
neutrinos  1.5 TeV. The achievable limit for
$M_{W_R}$ is now about 1.5 TeV at the triplet Higgs
resonance and outside the resonance  about 1 TeV,
 a considerable improvement to the present bound.
As the cross section is proportional to the mass of neutrino, the larger
$m_{\nu_2}$ the more stringent are the ensuing constraints.
Following the arguments of \cite{redbook}
we apply the following cuts:

\begin{itemize}
\item[--] Each b-jet should have energy more than 10 GeV.
\item[--] Each t-jet should have 
energy more than 190 GeV.
\item[--] The opening 
angle between two detected jets should be
greater than $20^{\circ}$.
\item[--]The angle between each detected jet and
the colliding axis should be greater than 
$36^{\circ}.$
\item[--]The total energy of the event should be
greater than 400 GeV.
\end{itemize}

\begin{figure}
\begin{center}
\epsfysize=10cm
\epsffile{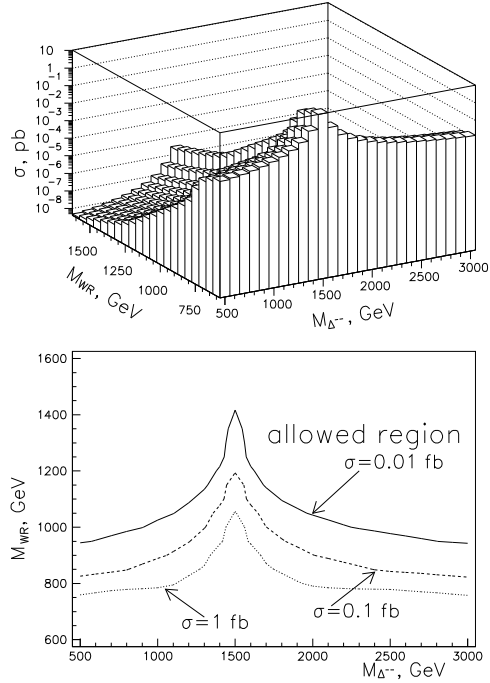}
\end{center}
 \caption{Cross section 
for the $e^-e^- \rightarrow b,b,\bar{t}, \bar{t}$
and it's contourlevels
at $\sigma=0.01 \: $ fb
(30 events per year),
 $\sigma=0.1 \: $ fb
(300 events per year), 
 $\sigma=1 \: $ fb
(3000 events per year)  
for the energy $E=1.5 \: TeV $, 
and the right-handed neutrino mass
$m_{\nu_2}=1.5 \: $ TeV. }
\label{s15n15}

\end{figure}

If we impose for the counterparts of the top 
quarks, the $c$ quarks, in the reaction $e^-e^- \rightarrow s\:s \:\bar{c}\: \bar{c}$ the  cut
$E_{\bar{c}_{1,2}} > 190$ GeV,
 which is very effective in diminishing 
 the SM background (see below), 
the cross sections differ not more than 12 \%.
Hence, one can immediately write down the following approximative relations  
 between the cross sections of the reactions with no, one and two $b$-jets in the final state:   
   
   \be
   \sigma(0b)\approx\sigma(1b)\approx 4 \cdot \sigma(2b); \:
   \lbl{rel}
   \ee
This relation may be very useful as a test
of the LR-model.

The main SM background of the reaction $e^-e^- \rightarrow b\:b \:\bar{t}\: \bar{t}$ is 
due to the process $e^-e^- \rightarrow  \nu_e \nu_e
b\:b \:\bar{t}\: \bar{t}$, which has the same visible particles in the final state. 
 If we impose the cut of 50 GeV on the energies of the final state electrons,
the cross section 
$\sigma(e^-e^- \rightarrow e^-\:e^-\:W^+\:W^-)$
diminishes by 3 orders of magnitude
and yields the background 
 at 1.5 TeV collision energy on the $0.03$ fb level.
There is a further suppression in the case
 of  the $b b \bar{t} \bar{t}$
due to the fact the intermediate $W_L$
bosons should actually be  away from the pole
as  the invariant mass of its decay products  $b,\bar{t}$
 should be greater than $m_t$.
This yields alltogether  8 orders of magnitude
    suppression  of the background, making it fully harmless.

\section{Summary}

It is shown that the reaction $e^-e^- \rightarrow q\:q\:\bar{q}\: \bar{q}$
may be observed at NLC for a wide range of reasonable parameter values of the
left-right symmetric model and already
below the $W_R $ threshold.
For the collision energy  $\sqrt s=1.5$ TeV
and luminosity $ 10 ^{35} {\rm cm}^{-2}
\cdot {\rm s}^{-1}$ the lower limit for the mass of
the right-handed gauge boson one could reach is $M_{W_R} \gsim 1000$ GeV.
Near the doubly charged Higgs ($\Delta^{--}$) resonance the lower bound on $M_{W_R}$ 
may reach, and even exceed, the value of the collision energy.
As the lepton number violation and neutrino masses are intimately connected through the Maojaran mass terms, the strength of 
the $e^-e^- \rightarrow q\:q\:\bar{q}\: \bar{q}$
process increases with the growth of the    mass of the right-handed neutrino.
The "non-diagonal" processes, i.e. the reactions where the $\bar q q$ pair or pairs in the final state mix with fermion families, 
are essentially suppressed, while all the "diagonal"
processes have approximately the same probability.
 Process $e^-e^- 
\rightarrow b\:b\: \bar{t}\: \bar{t}$
can be identified as $b$-tagging is possible.
For the processes involving only light quarks
 or containing just one $b$-jet are
approximately  related
to this    cross section by eq. (2).

The SM background can be suppressed
to the level 4 orders of magnitude below the
process rate if the proper
cuts in the phase space are
applied, and it can be made even 7 orders of magnitude below the signal level    if the full energy of the event can be
reconstructed with the  accuracy of 50 GeV.

\newcommand{\bi}{\bibitem}

\end{document}